\documentstyle[prd,aps,floats,epsfig]{revtex}

\begin{document}
\preprint{astro-ph/0111605} \draft

\input epsf

\newcommand{\yy}{$y_{i}$}
\newcommand{\dd}{$\delta_{NL}(\mathbf{x})$}
\newcommand{\be}{\begin{equation}}
\newcommand{\ee}{\end{equation}}
\newcommand{\ba}{\begin{eqnarray}}
\newcommand{\ea}{\end{eqnarray}}
\newcommand{\bet}{\mbox{\boldmath $\eta$}}
\newcommand{\bolk}{{\bf k}}
\newcommand{\bolr}{{\bf r}}
\newcommand{\pd}{\partial}
\newcommand{\fg}{{\mathcal G}}
\newcommand{\nn}{\nonumber \\}
\newcommand{\nnb}{\begin{displaymath}}
\newcommand{\nne}{\end{displaymath}}
\newcommand{\de}{\partial}
\newcommand{\nbar}{\bar{n}}
\newcommand{\k}{\mbox{\boldmath $k$}}
\newcommand{\x}{\mbox{\boldmath $x$}}
\newcommand{\r}{\mbox{\boldmath $r$}}
\newcommand{\rhatb}{\mbox{\boldmath $\hat{r}$}}
\newcommand{\y}{\mbox{\boldmath $y$}}
\newcommand{\s}{\mbox{\boldmath $s$}}
\newcommand{\g}{\mbox{\boldmath $g$}}
\newcommand{\E}{\mbox{\boldmath $E$}}
\newcommand{\C}{\mbox{\boldmath $C$}}
\newcommand{\nablab}{\mbox{\boldmath $\nabla$}}
\newcommand{\hMpc}{\,h^{-1}{\rm Mpc}}
\newcommand{\e}{\varepsilon}
\newcommand{\lm}{{\ell m}}
\newcommand{\lmd}{{\ell' m'}}
\newcommand{\sqtpi}{\sqrt{\frac{2}{\pi}}}
\newcommand{\rhat}{\hat{\r}}
\newcommand{\khat}{\hat{\k}}
\newcommand{\rh}{\hat{r}}
\newcommand{\Mpc}{\, h^{-1}{\rm Mpc}}
\newcommand{\sqdeg}{\,{\rm sq.}\, {\rm deg.}}
\newcommand{\eF}{{\cal F}}

\newcommand{\etab}{\mbox{\boldmath $\eta$}}
\newcommand{\rgl}{\rangle}
\newcommand{\lgl}{\langle}
\newcommand{\sky}{{\rm sky}}
\newcommand{\rd}{{\rm r}'}

\twocolumn[\hsize\textwidth\columnwidth\hsize\csname
@twocolumnfalse\endcsname


\title{Imaging the 3-D cosmological mass distribution
with weak gravitational lensing}
\author{A.~N.~Taylor }
\address{Institute for Astronomy, Royal Observatory,
Blackford Hill, Edinburgh, EH9 3HJ, U.~K.}
\date{\today}
\maketitle


\begin{abstract}
 I show how weak gravitational lensing can be used to image the
3-D mass distribution in the Universe. An inverse relation to the
lensing equation, relating the lensing potential evaluated at each
source to the full 3-D Newtonian potential, is derived. I consider
the normal modes of the lensing problem and clarify the equations
using a small-angle approximation. Finally I consider the
prospects of using this method to estimate the 3-D matter
distribution from a realistic galaxy lensing survey.
\end{abstract}

 \pacs{PACS number(s): 04.20.-q, 98.80.-k, 98.80.Es
 \hfill astro-ph/0111605}

\vskip2pc]


\section{Introduction}

Gravitational lensing provides us with the most direct and
cleanest of methods for probing the distribution of matter in the
universe \cite{mellier,schneiderBart}. The lensing effect arises
from the deflection of light by perturbations in the metric. These
deflections stretch and contract bundles of light rays, causing
the distortion of background galaxy images. Hence gravitational
lensing does not depend on any assumptions about the state of
matter. These distortions manifest themselves as a shear
distortion of the source galaxy image \cite{tysoncluster,ks}, or a
change in the surface number density of source galaxies due to
magnification \cite{btp,fort,taylor98} and can be used to map the
two dimensional projected matter distribution of cosmological
structure. As the matter content of the universe is dominated by
non-baryonic and non-luminous matter, gravitational lensing is the
most accurate method for probing the distribution of this Dark
Matter. Imaging of the Dark Matter distribution is a vital key to
understanding its nature.

 The measurement of the gravitational mass distribution in clusters
 of galaxies using gravitational lensing shear and
 magnification effects is now a well established
 technique, while on larger scales the detection of a cosmic
 shear signal \cite{cosmicshear} shows that
 the cosmological matter distribution can
 also be probed this way. A problem with these methods
 is the limited use of depth information, such as spectroscopic or
 photometric redshifts. Usually the
 intervening lensing matter distribution is approximated by a
 sheet. Some depth information can be introduced by
 lens tomography \cite{seljak98,Hu2000}, where the background source galaxies
 are divided into bins and the matter distribution can
 be approximated by a series of sheets. What is lacking is a
 method for estimating the full 3-D matter
 distribution from gravitational lensing. Here I address this problem.

\section{Method}

 The metric of a perturbed Friedmann-Lema\^{i}tre-Robertson-Walker
 universe in the conformal Newtonian, or longitudinal, gauge is
 \be
    ds^2 = - (1+2 \Phi)dt^2 + a^2(t) (1-2 \Phi) dr_i dr^i
 \ee
 where $\Phi$ is the Newtonian potential, $a$ is the cosmological
 scale factor, and we have assumed a
 spatially flat universe for simplicity. The Newtonian potential
 is related to the matter density field by Poisson's
 equation
  \be
    \nabla^2 \Phi = 4 \pi G \rho_m \delta a^2 = \frac{3}{2}
    \lambda_{H}^{-2} \Omega_m a^{-1} \delta,
  \ee
  where $\lambda_H=1/H_0 \approx 3000 \Mpc$ is the Hubble length, and
  $\Omega_m$ is the present-day mass-density parameter.

The lensing potential, $\phi$, for a source in a spatially flat
universe at distance $r$ is given by \cite{schneiderBart,HuSph}
 \be
    \phi(\r) =  2 \int_0^{r}\!\! dr' \,
     \left(\frac{r-r'}{r r'}\right) \Phi(\r'),
 \label{pot}
 \ee
 measured
 at an angular position $\rhat$ on the sky. We have assumed
 the Born approximation, where the light path is
 unperturbed. This equation shows that the
 lensing potential is a radial projection of the 3-D
  gravitational potential, with a radial Greens function
  $G(r,r')=2(r-r')/rr' \Theta(r-r')$, where $\Theta(r)$ is the Heaviside
  function.


The inverse relation to equation (\ref{pot}) is
 \be
    \Phi(\r) = \frac{1}{2} \de_r r^2 \de_r \, \phi(\r)
 \label{inv}
 \ee
 where $\de_r = \rhat.\nabla$ is the radial derivative, and we
assume the lensing potential has been appropriately smoothed to
allow differentiation. This can be verified by substitution into
equation (\ref{pot}) and integrating by parts.

The lensing potential is not an observable. The observables are
the dimensionless, symmetric, tracefree shear matrix,
$\gamma_{ij}$, which describes the distortion of the lensed image
and the magnification, $\mu$, which describes the change in area.
The shear metric is
 \be
    \gamma_{ij} = \left(\de_i \de_j - \frac{1}{2} \delta^K_{ij}
    \de^2\right)\phi,
 \ee
 where $\de_i \equiv r (\delta_{ij}- \rh_i \rh_j) \nabla_j =
r(\nabla_i - \rh_i \de_r)$ is a dimensionless, transverse
differential operator, and $\de^2 \equiv \de_i \de^i$ is the
transverse Laplacian. The indices $(i,j)=(1,2)$. In this
expression we have assumed a flat sky. The magnification is given
by \cite{btp}
 \be
    \mu = |(1-\kappa)^2 - \gamma^2|^{-1} \approx 1+ 2 \kappa.
 \ee
 The second approximation holds for weak magnification, where the lens
convergence, $\kappa$, is defined by the 2-D Poisson's equation,
 \be
    \kappa = \frac{1}{2} \de^2 \phi.
 \ee
 In principle there is another second-rank tensor which can be formed
 from a scalar potential;
 \be
    B^{ij} = \varepsilon^i_{(m} \de_{n)} \de^j \phi_B,
 \ee
 where $\varepsilon^i_j$ is the 2-D antisymmetric Levi-Civita tensor, the
 brackets indicate symmetrization of the indices and $\phi_B$ is a
 pseudo-scalar potential. The symmetric
 tensor $B$, and pseudo-scalar $\phi_B$ have odd parity and therefore
 cannot correspond to the
 parity invariant matter density field. Hence we expect $B=0=\phi_B$, and
 can use this to investigate noise and boundary
 effects in lensing.

 The lensing potential can be estimated from the shear field by the
 generalised Kaiser-Squires \cite{ks} relation;
 \be
        \tilde{\phi} = 2  \de^{-4} \de_i \de_j \ \gamma_{ij},
 \label{kaiser-squires}
 \ee
 where $\de^{-2}$ is the inverse 2-D Laplacian operator.
 In practice the shear field is only discretely sampled by
 galaxies so we must smooth the shear field to perform the
 differentiation. This also serves to make the uncertainty on the
 measured shear field finite, since each source has an unknown
 intrinsic ellipticity. This smoothing need only be perpendicular to the
 light path. There may also be intrinsic alignments of
 galaxies due to tidal effects during their formation
 \cite{Hoyle,alignments}. However,
 these appear to be small at large distances \cite{brown}.

The observable shear and convergence allow us to measure the
lensing potential up to an arbitrary function of $r$, the radial
distance; \be
    \tilde{\phi}(\r) = \phi(\r) + \psi(r),
\ee where $\tilde{\phi}$ is the measured gravitational potential
and  $\psi(r)$ is a solution to the equations
 \be
    \left(\de_i \de_j - \frac{1}{2} \delta^K_{ij}
    \de^2\right)\psi =0 .
 \ee
 This sheet-like gauge freedom arises because at each distance the shear and
 convergence define the potential only up to an arbitrary constant. In
 principle this can arbitrarily change as a function of
 distance. As the
 reconstruction of the Newtonian potential requires radial
 derivatives we have to smooth in the
 radial direction. This also transforms $\psi$
 from an arbitrary radial function to one which is smooth
 on the scale of the radial smoothing radius. The radial gauge freedom we see
 here is related to the so-called sheet-mass degeneracy which
 also arises as a constant of integration when deriving the
 convergence from the
 shear \cite{Falco}. However it is important to note that the radial
 freedom we see here also arises if the convergence is used
 to estimate the lensing potential.

Since we expect the true lensing potential field to respect
statistical isotropy and homogeneity, the observed
$\tilde{\phi}$-field must obey the equation
 \be
    \lgl \de_r \tilde{\phi} \rgl = \frac{\de }{\de r} \psi,
 \ee
 where the angled brackets $\lgl \cdots \rgl$ denote ensemble
 averaging, or
 tangential averages over light-paths at the same distance.
Substituting back into equation (\ref{inv}) we find
 \be
    \Phi =  \frac{1}{2} \de_r r^2
    ( \de_r \tilde{\phi} -\lgl \de_r \tilde{\phi}
    \rgl)
    \label{unbiaspot}
 \ee
 is an unbiased estimate of the Newtonian gravitational potential.
 The derivation of equation (\ref{unbiaspot}) is the main result
 of this paper, and demonstrates that the full 3-D
 Newtonian potential, and hence the matter-density field, can be
 reconstructed from weak lensing observations.

 In practice $\psi$ will not be a major problem for
 large surveys, since the boundary conditions used for the inversion
 can be used to set the mean potential at a given radius to zero.
 This is fine if the survey is large enough and the mean potential
 is zero, but if the angular size of the survey is small
 the potential may not average to zero.

\subsection{Normal modes}

The lensing functions can conveniently be expanded in normal
modes, this time taking the curvature of the sky into account. We
define the 3-D spherical harmonic modes of a field by
 \be
    \phi_\lm(k) = \sqtpi \int_0^{r_\infty}\!dr r^2
    \int_{4 \pi}\!\! d\Omega \, \phi(\r) j_\ell(k r) Y_\lm^*(\Omega),
 \ee
 where $r_\infty=\int_0^\infty dt/a(t)$ is the causal horizon.
 For a spatially open universe with hyperbolic geometry the usual
 spherical Bessel functions can be generalised to
 the hyper-spherical Bessel functions, $j_\ell(x) \rightarrow
 X_\ell(\Omega_K,x)$.

 The harmonic moments of the potential are related to the
 convergence by a 2-D Poisson's equation, yielding \cite{HuSph}
 \be
    \kappa_\lm(k) = -\frac{1}{2} \ell(\ell+1) \phi_\lm(k).
 \ee
 Similarly the shear field can be decomposed into tensor
 spherical harmonics (e.g. \cite{stebbins}),
 \be
    \gamma_{ij}(\rhat)  =
    \sqtpi \int_0^\infty \! \! k^2 dk\,
    \sum_\lm \gamma_{\lm}(k) j_\ell(k r) Y^{E}_{(\lm) ij}(\rhat),
 \ee
 where
 \be
    Y^{E}_{(\lm) ij}(\rhat) = \sqrt{\frac{2(\ell-2)!}{(\ell+2)!}}
    (Y_{(\lm);ij}-1/2 g_{ij} Y_{(\lm);c}^{\hspace{0.7cm} c})
 \ee
 and where $g_{ij}$ and $``;"$ are
 the metric and covariant derivative on the 2-sphere respectively.
 The $\gamma_{\lm}(k)$ are then related to the lensing potential
 field by \cite{HuSph}
 \be
    \gamma_{\ell m}(k) = \frac{1}{2} \sqrt{\frac{(\ell+2)!}{(\ell-2)!}}
    \phi_{\lm}(k).
 \ee
 Similar expressions exist for the non-gravitational
 parity-violating shear term.

 The normal modes of the Newtonian potential are then related to
 the lensing potential by
 \be
    \Phi_\lm(k) = \frac{1}{2} \int\! dk' \, [ \ell(\ell+1)
    \delta_D(k'-k)    - \alpha_\ell(k,k') ] \phi_{\lm} (k')
 \label{modes}
 \ee
 where
 \be
        \alpha_\ell (k,k') = \frac{\pi}{2}
        \int_0^\infty \! dr \, (k'r)^4 j_\ell(k r)
        j_\ell(k' r).
 \ee
 The first term of equation (\ref{modes}) is simply minus the lens
 convergence, which reflects the angular
 distribution of the matter field. The second term contains
 all of the distance information and shows that the radial modes are
 correlated, as lensing accumulates with distance.

\subsection{The small-angle approximation}


 For small-angles these equations can be simplified.
 It is useful to consider plane-wave solutions where
 the inversion equation becomes
 \be
    \Phi = - \frac{1}{2} r^2 k^2 \mu^2 \phi,
 \ee
 and now $\mu=\khat . \rhat$ is a cosine angle. Combining this with the
 plane-wave solution for the convergence,
 \be
    \kappa = - \frac{1}{2} r^2 k^2 (1-\mu^2) \phi,
 \ee
 and Poisson's equation we find
 \be
    \delta = - \frac{2}{3} k^2 \lambda_H^2
        \left( \frac{a}{\Omega_m}\right)
        \left(\frac{\mu^2}{1-\mu^2} \right) \kappa.
        \label{smallrec}
 \ee
 This relation can be readily understood if we consider modes
 parallel and perpendicular to the light-path, $k_{||}=k\mu$ and $k_{\perp}
 = k (1-\mu^2)^{1/2}$ respectively. While $k_\perp$ is related to the
 tangential scale of the structure being probed, the radial mode,
 $k_{||}$ arises from the integration along the light-path and
 so shows lensing probes structures along the light-path with
 wavenumber
 $k_{||}\sim 1/r$, where $r$ is the distance to the source. If we fix
 the sources at a single distance from the observer, the lensing
 inversion equation reduces to the 2-D Kaiser-Squires relation
 \be
        \kappa = - \frac{3}{2} \frac{r^2}{\lambda_H^2}
        \left( \frac{\Omega_m}{a} \right)
        \left( \frac{k_\perp^2 r^2 }{1+ k_\perp^2 r^2} \right)
        \delta(k_{||}=r^{-1},k_\perp)
 \ee
 and so for large distances or small structure,
 we find that the lensing signal increases as a function of
 the source distance as $\kappa \sim r^2$.

\subsection{Measurement of the mass-density field}

 These relations allow us to estimate the accuracy for
 reconstructing the 3-D density field from gravitational lensing.
 The covariance matrix of Fourier modes of the density field,
 $\delta(\k)$, estimated from lensing is
 \be
    C_{\delta \delta}(\k) = P_\delta(k) + \frac{4}{9}
    \left(\frac{ak^2 \lambda_H^2\mu^2}{\Omega_m(1-\mu^2)}\right)^2
    \frac{e_{\rm rms}^2}{n},
 \ee
 where we have assumed the underlying density field is
 statistically homogeneous and isotropic;
 \be
    \lgl \delta(\k) \delta^*(\k')\rgl = (2 \pi)^3 P(k)
    \delta_D(\k-\k'),
    \label{noise}
 \ee
 and $\delta_D(\k)$ is the Dirac delta function.
 The second term arises due to Poisson sampling of the shear field
 by source galaxies, where $e_{\rm rms}\approx 0.4$ is the intrinsic
 dispersion of galaxy ellipticities, and $n$ is the density
 of sources.

 The uncertainty on a measurement of the matter power spectrum
 from a finite galaxy lensing survey is
  \be
    \frac{\Delta P_\delta(k)}{P_\delta(k)} =
        \frac{2 \pi}{\sqrt{k^3 d \ln k V_{\rm eff}}},
    \label{fish}
   \ee
   where $V_{\rm eff}=\int d^3r  [P(k)/(P(k)+N(\k))]^2$ is the
   effective volume of the survey, where $N(\k)$ is the Poisson noise,
   and we sample the matter power spectrum
   in logarithmic intervals of $d\ln k$. Equation (\ref{fish})
   shows the fractional uncertainty per modes divided
   by the square root of the number of effective independent modes that will
   fit into the survey volume. For a finite survey the angular
   part of the integral should have limits $1/(kR) \le \mu \le
   \sqrt{1-1/(k R)^2}$ and $k\ge \sqrt{2}/R$, where $R$ is the
   size of the survey.
   For small scales this is dominated by
   shot-noise and $k^3 V_{\rm eff}(k)/2 \pi^2 \approx 27/(56 \pi)
   (nP/e_{\rm rms}^2)^2 (k \lambda_H)^2 (R/\lambda_H)^{10}(\Omega_m/a)^4$,
   which is strong function of the volume.

   Figure 1 shows the 3-D matter power spectrum and the expected
   uncertainty from an idealised galaxy lensing survey. The dotted
   line is the linear power spectrum, while the solid line is the
   nonlinear power \cite{pd}.
   The model
   is a $\Lambda$CDM model, with $\Omega_\Lambda=0.7$ and
   $\Omega_m=0.3$, normalised to the present-day cluster abundance
   \cite{whiteetal}. The upper dot-dashed line shows the
   uncertainty for a survey with source density $n=10^{-3} \Mpc^{-3}$
   and volume $V=10^9
   \Mpc^3$, similar to the Sloan Digital Sky Survey. The lower
   lines are for larger surveys with the lower line for a survey
   ten time larger. Clearly the 3-D matter power
   spectrum can be recovered to good accuracy.

\begin{figure}
\centering
\begin{picture}(200,200)
\includegraphics{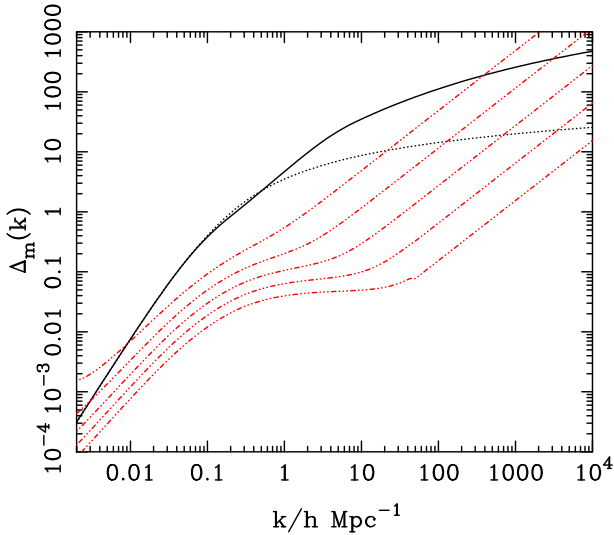}
\end{picture}
\caption{The dimensionless matter power spectrum,
$\Delta_m(k)=\sqrt{k^3P(k)/2\pi^2}$, for a $\Lambda$CDM cosmology,
with $\Omega_\Lambda=0.7$ and $\Omega_m=0.3$, normalised to the
observed abundance of clusters. The dotted line is linear theory,
the solid line is nonlinear theory. The dot-dashed lines are the
expected uncertainty on a measurement of the power from a set of
large-scale galaxy lensing surveys.} \label{fig4}
\end{figure}

\section{Summary}

 I have derived an exact expression for the reconstruction of the
 full three-dimensional matter density field from measurements of
 weak gravitational lensing of source galaxies. This can be
 applied to matter fields in both the linear and nonlinear
 clustering regime, since the method is based on a new relation
 between the Newtonian potential and the gravitational
 lensing potential. The lensing potential can be estimated from
 observable lens shear or convergence fields. A new sheet-like
 degeneracy arises due to a constant of integration of the
 observable fields which is an arbitrary function of distance. I
 have shown that this function can be removed by averaging over light
 paths.


 I have derived the normal modes of the lens fields and the
 reconstruction equation. Using a small-angle
 and plane-wave approximation I have derived a simplified set of
 equations. The small-angle approximation allows us to estimate
 the accuracy to which a realistic galaxy lensing survey can
 reconstruct the full matter distribution. I find that the
 reconstruction of the full distribution is possible with current
 surveys, if the data quality is sufficiently good. 3-D imaging
 is an interesting challenge for future
 lensing surveys. Since the addition of depth information
 adds an important new aspect to lensing, estimation of
 statistical quantities such as the 3-D matter power spectrum
 should be possible with near-future surveys. Combined with the
 distribution of galaxies from the same survey, this will lead to
 direct constraints on the environmental aspects of galaxy
 formation, as well as a direct probe of the cosmological
 distribution of Dark Matter, which in turn will place strong constraints
 on theories of structure formation and the nature of the Dark Matter.

\section*{Acknowledgements}

ANT is supported by a PPARC Advanced Fellowship, and thanks Alan
Heavens, David Bacon and Simon Dye for useful discussion while
this work was in progress.


\end{document}